# Early work on pion-nucleon scattering at Rochester

G. Giacomelli
University of Bologna and INFN Sezione di Bologna
giacomelli@bo.infn.it

**Abstract.** In 1956 when I was a graduate student at the University of Rochester, the post doc Herman Winick arrived there, joined the group and we worked together with colleagues from different countries on low energy pion-proton elastic scattering at the cyclotron. The small experiment was highly successful and was the beginning of a long research life and of friendship.

## 1. Introduction

In the late '40s Sidney Barnes and several colleagues built the 130-inch Rochester cyclotron [1]. In the US a number of small cyclotrons, together with some relatively small nuclear emulsion laboratories of cosmic rays in the US and in Europe, opened the way to experimental high energy nuclear and particle physics. The energy of the 130-inch cyclotron at Rochester was about 240 MeV, below the threshold for pion production in p-p collisions. Nevertheless pions were abundantly produced because of the Fermi motion of the nucleons inside the heavier target nuclei such as carbon. With these pions, many early experimental investigations were performed on pi mesons, in particular on low energy $\pi$–proton scattering.

The pion-proton scattering was considered fundamental to understand classical nuclear physics. Thus differential and total cross sections were measured at ever increasing energies. The data were analyzed in terms of phase shift analyses and it became soon obvious that one had to select the Fermi type set of phase shifts with a dominant $\alpha_{33}$ which accounted for the first $\Delta(1232$ MeV$)$ resonance.

Soon, after the end of world war 2 in the late '40s and '50s, Robert Marshack started to attract to Rochester many foreign PhD students from Japan, Italy and other countries, and later he started the Rochester series of High Energy Nuclear and Particle Physics Conferences, which became an international important recurrent event. In that period the atmosphere in the Department of Physics and Astronomy of the University of Rochester was very stimulating with regular courses and seminars and many discussions (I remember very well the beautiful seminar on parity violation by C. N. Yang). The majority of foreign students came from Japan and Italy, two of the nations which had lost world war 2. The friendship between Japanese and Italians started immediately and lasted also after most of the graduate students went back to their homelands.

In the fifties the Kodak company, with its headquarter in Rochester, was doing very well and was helping to finance the University. Some colleagues after visiting the leaning tower in Pisa, used to say that the many photographs of its tower, made by so many tourists, was effectively giving one of the most important financial contributions to the University of Rochester !

With a Fellowship from the University of Rochester and a Fulbright travel grant, I started my graduate studies at Rochester in 1954 and I was one of the few italians to obtain a PhD in Physics, in 1958. This degree was useless in Italy, but was important in my later visits to the US, to Brookhaven, Fermilab and the Universities of California at Berkeley and at Riverside.

## 2. Elastic pion-nucleon scattering at the 130 inch cyclotron

In Rochester I started research activities at the cyclotron in the group led by Sidney Barnes, and I worked with some young american students, (Jim Ring and Doug Miller), a British researcher from Harwell (Basil Rose), a Japanese graduate student ( Kozo Miyake from Kyoto) and for slightly more than one year with Herman Winick, who was a post-doc from Columbia University. With all of them, the friendship started in Rochester, continued forever.

Fig. 1 shows the layout of the cyclotron, of the shielding and of the 41 MeV negative pion beam. The pions emerged from the cyclotron through a thin window and traveled in a helium path through an opening in a 3-feet brass shield into a magnet which deflected and focused the beam that passed through another opening in a lead shielding, to the approximate focus. The pions then came into open air and were counted by a telescope made of two small disk-shaped scintillation counters, 1 and 2 in Fig. 2. The available pion current was about $1-3 \times 10^4$ pions per minute.

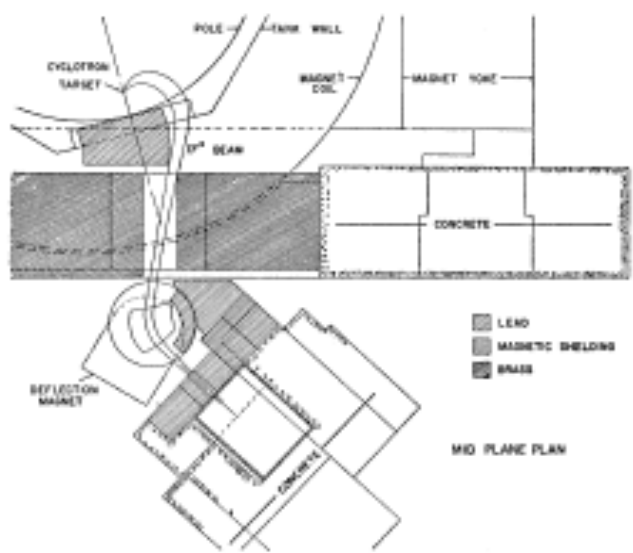

Fig. 1. Layout of the Rochester cyclotron and the shielding for the 41 MeV meson beam.

The tipical layout of the elastic scattering experiment for scattering angles around 90° is shown in Fig. 2; it involved the liquid hydrogen target, which had an ample reservoir; its shape was cylindrical and the beam came along its axis. The cylinders were made of stainless steel foils silver-soldered to shape. The inner cylinder was of 0.002-in foil, while the outer one was 0.005-in. The separation between the two cylinders was 3/16-in. The end windows were of 0.002-in beryllium-copper foils. (This configuration shown in Fig. 2 was called the "no wall scattering" configuration). Counters 1 and 2 defined the beam; checks were made to verify beam parallelism. Counters 3 and 4 defined the acceptance of the scattered pion and the solid angle. As a matter of fact, counter 4 was slightly smaller and better defined the acceptance. Counter 5 was used as an anticoincidence to reduce the random background. The trigger was $123 4\bar{5}$; it opened a gate which allowed the signal from counter 3 to be sent to a 20 channel pulse-height analyzer.

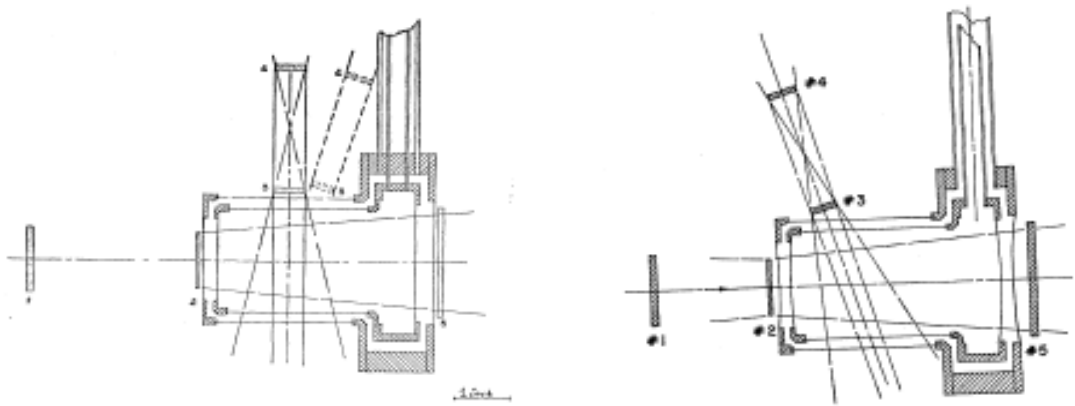

Fig. 2a. left. Geometry used for the 90° and 72.5° scattering experiments. 2b right. Target-counter arrangement for the 107° measurement.

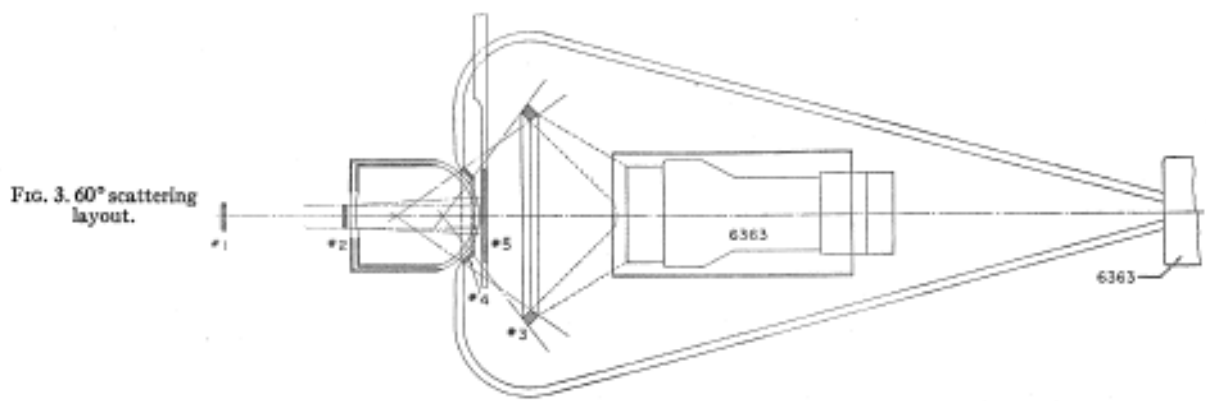

Fig. 3. 60° scattering layout.

The geometry described in Fig. 2 was particularly simple and was used to measure elastic scattering at 72.5°, 90° and 107.5°. For scattering at smaller (60°) and at larger angles (150°) the geometry was somewhat more complicated, as illustrated in Fig. 3 for scattering at 60°.

The energy of the pion beams was determined from range measurements.

The effective contamination by muons and electrons depended on the gain settings of the counters 1 and 2 in the incident telescope. The contamination was well measured from the pulse height distributions obtained from counter 1 and counter 2 set at high voltages, (which meant higher gains) and also when used at the normal gains set for running the experiment. The peaks arising from electrons and muons were well separated.

In the negative beam the effective muon contamination was about 3%. In the positive beam the muon-positron contamination was about 6%. The proton contamination was easily removed.

The ratio of the counting rates with the full target and the empty target was measured and a correction was made. Small corrections were also made for counting losses and for the finite angular resolution of the apparatus.

## 3. Analyses and Results

Phase shift analyses were made using an IBM650 computer, which was the first available computer used systematically in physics. The used formulae needed approximations at low energies, particularly to obtain the low energy parameters of pion-nucleon scattering. The dominant phase shift was the isotopic spin 3/2 $P_{33}$ phase shift $\alpha_{33}$, which has a well known large resonance at $m_\Delta=1232$ MeV. The isotopic spin ½ phase shift $\alpha_{13}$ was well determined as function of momentum, see Fig. 4. The S-wave phase shifts, which were considered as important low energy parameters, were computed and compared with several models of low energy parameters, see Fig. 5. The experimental differential cross sections measured by higher energy experiments for both $\pi^+p$ and $\pi^-p$ were compared with smooth extrapolated curves obtained from the Rochester lower energy measurements: there was a good agreement , which could be considered a good check for the low energy measurements and of the used methods of phase shift analyses,  see Fig. 6.

Good agreement was also found between $\pi^-p$ forward scattering amplitudes and a dispersion equation in the energy region discussed here.

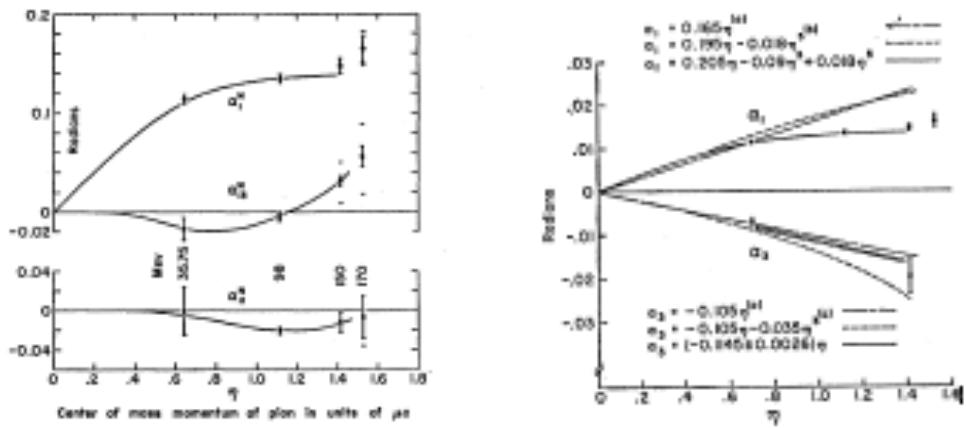

Fig. 4 left:T=1/2 phase shifts as a function of momentum.  Fig. 5 right: S-wave phase shifts compared to various prescriptions [J. Orear, Nuovo cimento 4, 856 (1956); (b) H. L. Anderson, Proceedings of the Sixth Annual Rochester Conference on High-Energy Nuclear Physics, 1956 (Interscience Publishers, New York, 1956); (c) Ferrari et al. , CERN 1956.

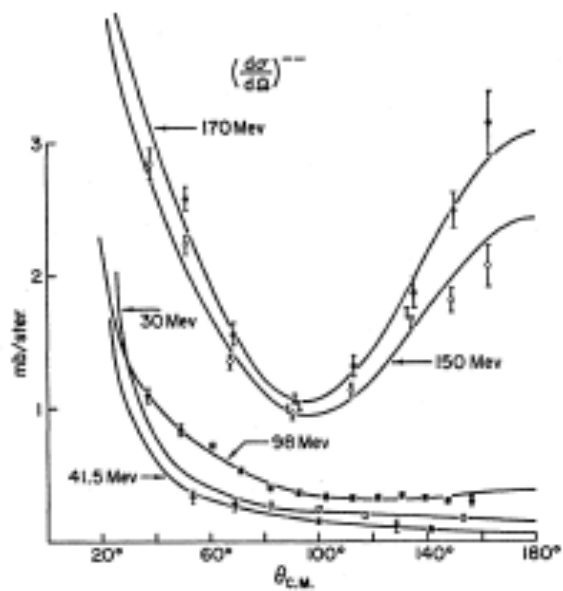

Fig. 6 : Experimental differential cross sections with smooth curves obtained by the Rochester phase shift analysis.

## 4. Conclusions

The early high energy nuclear experiments were simple experiments performed by small groups of physicists with a small number of scintillation counters (5 counters at Rochester, 6-7 counters in Columbia [5]).

Later experiments started to be more complex, as, for example, at CERN with a large number of spark chambers [6] and at Harvard [7]. The complexity, the number of electronic channels and the number of physicists and engineers increased incredibily in the large experiments at Fermilab, Brookhaven and now at the CERN LHC .

In the early experiments much equipment was built by graduate students and post docs [8] and most of the analyses programs were made by a few graduate students [9]. Now most experiments require large numbers of physicists, engineers and specialized programmers.

It is interesting to note that the few colleagues working in the small experiment at Rochester kept strong contacts with each other and met often in different places for pleasure, and for work (in England at Harwell [10], in Japan and in CERN [11,12]) and for seminars and colloquia.

I add some recollections of the social life in Rochester from Pina, who became my wife in 1958, and a later episode from Bologna.

-------------------------------------------------------------------------------

In September 1956, thanks to a fellowship from the American Association of University Women (A.A.U.W.) and a Fulbright travel grant, I ended up at the University of Rochester, N.Y. I had a degree in Biology from the University of Bologna, Italy, where the biology field started in 1951, when I entered at the University. For me the new field was a real attraction which grew year after year.

The information about the fellowship competition for women and the possibility to continue my scientific research in Rochester, came from Giorgio Giacomelli who, after his physics degree in Bologna, in 1954, had moved there with a graduate fellowship and a travel grant.

My life in Rochester was great. I lived with a very nice family in a small house, not far from the University, which gave me the opportunity to know students and families (the free housing was part of the fellowship).

A real friendship began and continued with Herman Winick's family. Renee Winick was adorable with me. She accepted me as a sister!.....her baby was wonderful and enjoyed my visits. Renee was very much interested in my country.......and often we talked about cooking....Once she asked me if I knew how to make "lasagne bolognesi". I admitted that I could, but I also tried to explain how long and difficult it was to prepare all the ingredients needed. I did not know how to translate "matterello per la sfoglia", a long and heavy wooden stick used for pulling the dough to a thin sheet,.....so I decided to go around supermarkets to get a "matterello", but there was nothing to do and even asking people I had no success. Finally, passing along a brooms area of a store, I had a brilliant idea.....: a broomstick!....That stick was long enough, but too thin and light to squeeze a large "sfoglia"…. Anyway I decided I could try. I took the most perfect one and went to pay for it..... they were used to sell brooms alone, but not handles alone….and they had no price for it.....well they gave me the handle without charge!

Early on a Sunday morning I went to the Winicks with the idea of making "lasagne". I started preparing "ragu" and while it was cooking I prepared "besciamella", spinaches and grated cheese, not a real "parmigiano…of course. Then, I took flour and eggs to start the dough. More problems: I needed a very large not-sticky table over which I could smooth down the dough.....Finally I found a possibility but it was too high for me......no problem....I climed on a little stool and from there I rolled the "sfoglia".

Well, after a full day work, I put my "lasagne" in the oven.....and an hour later we were all at the table and the lasagne were really appreciated by everybody !
(Pina Maltoni)

Many italian toll roads cross Bologna (Italy) in several directions. A few years ago, in the early evening, we were expecting Renee and Herman Winick at our home, but we received a call from Herman: he was in the wrong direction after entering in a toll road, had no possibility to get off, and he could not explain his situation to the attendants. I answered that I would join them by car and see what could be done. I reached the area but I had to find how to go inside without entering the toll road with my car. I parked my car and I walked in a small unpaved road in an open field to find a passage : it was dark but the sky was very clear… I do not remember how I found a passage in a green hedge bordering the toll road entrance… anyhow I managed to get very close to the Winicks, who were still discussing with the toll road attendants. I first asked Herman how he reached that place and then I discussed with the attendants: he had lost the ticket and traveling only one exit on the toll road, a large fare was demanded as if he originated at the furthest distance.

And also my situation was not clear : how did I enter by foot in that place reserved to cars? After a long discussion I managed to convince the attendants of the mistake made by the two americans and explained that I left my car outside the toll road in an earth road in the open field close by (because of the clear evening the car was easily visible). At the end, very nicely, they allowed the Winicks' car to exit through a service gate, asked me to promise not do it again…. and they would close more properly the passage in the green hedge.

It took some time to walk to my car, reach the Winicks and then proceed all together…but we made it !
(Giorgio Giacomelli)

-------------------------------------------------------------------------------------------------------------

**Acknowledgements.** I would like to acknowledge the very good cooperation by all the "old" colleagues at the University of Rochester in the years 1954-57 and in later years.